\documentclass[sigconf, nonacm, screen, pdfa]{acmart}
\usepackage{textcomp}
\usepackage{booktabs} 
\usepackage{amsmath}
\usepackage{float}
\usepackage{algorithm}
\usepackage{algpseudocode}
\usepackage{graphics} 
\usepackage{graphicx}
\usepackage{subfigure}
\usepackage{caption}
\usepackage[np,autolanguage]{numprint}
\usepackage{units}
\usepackage{fontawesome}
\usepackage{balance}
\usepackage{listings}
\usepackage{color}
\usepackage{makecell}
\usepackage{multirow}
\setcitestyle{numbers,sort}
\colorlet{linecol}{black}
\usepackage{comment}
\usepackage{forest}
\usepackage{adjustbox}
\usepackage{hhline}

\usetikzlibrary{arrows.meta,angles,quotes}
\tikzset{
  my rounded corners/.append style={rounded corners=2pt},
}
\makeatletter
\def\BState{\State\hskip-\ALG@thistlm}
\makeatother

\definecolor{dkgreen}{rgb}{0,0.6,0}
\definecolor{gray}{rgb}{0.5,0.5,0.5}
\definecolor{mauve}{rgb}{0.58,0,0.82}
\definecolor{dkred}{rgb}{1.0,0.01,0.24}

\newcommand{\mat}[1]{\ensuremath{\mathbf{#1}}}
\newcommand{\card}[1]{\lvert #1\rvert}

\newtheorem{example}{Example}
\newtheorem{definex}{Definition}

\renewcommand{\arraystretch}{1.5}

  {\list{}{\leftmargin=0.14in\rightmargin=0.14in}\item[]}%
  {\endlist}

\newcommand{\eat}[1]{}

\AtBeginDocument{%
  \providecommand\BibTeX{{%
    \normalfont B\kern-0.5em{\scshape i\kern-0.25em b}\kern-0.8em\TeX}}}

\begin{document}

\title[]
{CAMEO: Autocorrelation-Preserving Line Simplification for Lossy Time Series Compression}

\author{Carlos Enrique Muñiz-Cuza}\thanks{Half of the work was done while the first author was affiliated with Aalborg University}
\affiliation{\institution{TU Berlin\country{Germany}}}
\email{muniz.cuza@tu-berlin.de}

\author{Matthias Boehm}
\affiliation{\institution{TU Berlin\country{Germany}}}
\email{matthias.boehm@tu-berlin.de}

\author{Torben Bach Pedersen}
\affiliation{\institution{Aalborg University\country{Denmark}}}
\email{tbp@cs.aau.dk}
\renewcommand{\shortauthors}{}

\begin{abstract}
Time series data from a variety of sensors and IoT devices need effective compression to reduce storage and I/O bandwidth requirements. While most time series databases and systems rely on lossless compression, lossy techniques offer even greater space-saving with a small loss in precision. However, the unknown impact on downstream analytics applications requires a semi-manual trial-and-error exploration. We initiate work on lossy compression that provides guarantees on complex statistical features (which are strongly correlated with the accuracy of the downstream analytics). Specifically, we propose a new lossy compression method that provides guarantees on the autocorrelation and partial-autocorrelation functions (ACF/PACF) of a time series. Our method leverages line simplification techniques as well as incremental maintenance of aggregates, blocking, and parallelization strategies for effective and efficient compression. The results show that our method improves compression ratios by 2x on average and up to 54x on selected datasets, compared to previous lossy and lossless compression methods. Moreover, we maintain---and sometimes even improve---the forecasting accuracy by preserving the autocorrelation properties of the time series. Our framework is extensible to multivariate time series and other statistical features of the time series.   
\end{abstract}

\maketitle

\begingroup
\renewcommand\thefootnote{}\footnote{\noindent
This work is licensed under the Creative Commons by Attribution 4.0 International License (CC BY 4.0). Visit \url{http://creativecommons.org/licenses/by/4.0/} to view a copy of this license. For any use beyond those covered by this license, obtain permission by emailing the owner/author(s). \\
\textcopyright2025 Copyright held by the owner/author(s) \\
}\addtocounter{footnote}{-1}\endgroup

\section{Introduction}
\label{sec:introduction}

High-frequency time series are everywhere, from industrial manufacturing to weather prediction. For instance, an offshore oil rig typically has 30,000 sensors, of which only a few are utilized for real-time control and anomaly detection~\cite{manyika2015unlocking}. Time series compression can significantly reduce storage space, I/O bandwidth (storage or network) and analysis requirements~\cite{influxdata2023influxdb,Jensen2023,Zhang22CompressDB,jensen2017time,damme2017lightweight}. Motivated by these benefits, numerous algorithms have been proposed for lossless~\cite{pelkonen2015gorilla, liakos2022chimp, campobello2017rake,pope2018accelerometer,DBLP:journals/pvldb/XiaoHHSHW22} and lossy~\cite{eichinger2015time, marascu2014tristan,khelifati2019corad,chandak2020lfzip,blalock2018sprintz} compression. While lossless methods preserve the original data, lossy methods offer an appealing trade-off: more effective compression with only small, typically bounded reconstruction error.

\textbf{Lossy Compression Problem:} To reduce the impact on downstream applications, lossy compression often minimizes the reconstruction error. These methods focus on maximizing compression ratios, bounded to a maximum distortion of time series values~\cite{DBLP:journals/pvldb/ElmeleegyECAZ09,KitsiosLPK23,DBLP:conf/icde/LuoYCLFHM15,chandak2020lfzip,martinez2012more,tirupathi2022machine}.
Common techniques include domain transformation (Fourier Transform)~\cite{karim2011wavelet,cooley1965algorithm,ahmed1974discrete}, functional approximation (Polynomial Approximation)~\cite{eichinger2015time,keogh2001dimensionality,DBLP:conf/sigmod/KeoghCMP01,DBLP:journals/pvldb/ElmeleegyECAZ09,DBLP:conf/sigmod/NgC04}, or symbolic representation (Dictionary Encoding)~\cite{shieh2008sax,lin2007experiencing,marascu2014tristan,DBLP:conf/dmkd/LinKLC03,lin2007experiencing,krawczak2014approach}. Across these techniques, the reconstruction quality is typically computed by a single metric like Normalized Root Means Square Error (NRMSE) or the Peak Signal-to-Noise Ratio (PSNR)~\cite{tao2019z}. Despite bouding these errors, the resulting impact on different time series analytics remains largely unclear, requiring a tedious, semi-manual trial-and-error exploration~\cite{cappello2020fulfilling,ochoa2017effect,hollmig2017evaluation,moon2018evaluating,eichinger2015time,cuza2024evaluation}. 

\begin{figure}[t]
	\includegraphics[scale=0.57]{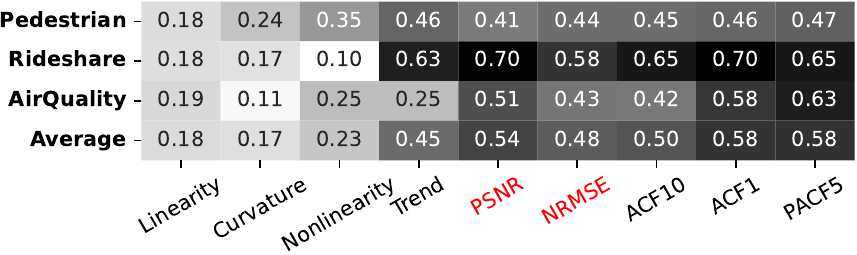}
	\vspace{-0.25cm}
	\caption{Pearson Correlation between the Impact on Forecasting Errors and Different Statistical Time-Series Features.}
	\label{fig:forecasting_corr}
\end{figure}

\textbf{A Case for Preserving Statistical Features:} Instead of limiting the reconstruction error, more complex statistical features can represent better downstream analytics, such as forecasting and anomaly detection. For example, we conduct an experiment using the Discrete Fourier Transform (DFT)~\cite{cooley1965algorithm} as a lossy compressor, the Seasonal-Trend decomposition using LOESS with Exponential Smoothing (STL-ETS) as a forecasting model~\cite{cleveland1990stl,hyndman2018forecasting}, and three datasets~\cite{DBLP:conf/nips/GodahewaBWHM21}. The three datasets, Pedestrian, Rideshare, and AirQuality, contain hourly sampled data from 66, \numprint{2495}, and 270 different sensors (\numprint{2831} time series in total). We measure the reconstruction error by computing the NRMSE and PSNR at each compression level. Moreover, we compute the impact on the forecasting accuracy (measured by the Modified Symmetric MAPE~\cite{hyndman2018forecasting}) and the impact on different time series features~\cite{R-tsfeatures} (including the ACF and PACF). Figure~\ref{fig:forecasting_corr} shows the Pearson correlation between forecasting errors and deviation of statistical features under a range of different compression levels. The results show that the features related to the series autocorrelation and partial autocorrelation functions---namely ACF1 and PACF5, have a stronger correlation to the impact on forecasting accuracy than NRMSE and PSNR. Intuitively, the ACF and PACF implicitly capture temporal dependencies (e.g., seasonality) relevant to the forecasting model. Therefore, a lossy compression technique that retains these features is more likely to maintain forecasting accuracy across different models. 

\textbf{Contributions:} In this paper, we introduce a new auto\-\textbf{c}or\-rel\textbf{a}tion-preserving lossy ti\textbf{m}e s\textbf{e}ries c\textbf{o}mpressor (CAMEO). The key objective is to guarantee a user-provided maximum deviation of the ACF or PACF on compressed data. CAMEO employs an iterative greedy approach to achieve this goal, eliminating points sorted by their impact on the ACF or PACF. The ACF and PACF statistics are updated incrementally by maintaining only a few aggregates. To improve runtime performance, we further leverage blocking and parallelization strategies. Our technical contributions are:

\begin{itemize}
	\item We survey lossy time series compression and line simplification through a new hierarchical classification in Section~\ref{sec:background}. 
	\item We introduce a new problem formulation for lossy time series compression that guarantees user-provided constraints with respect to specific statistical features in Section~\ref{sec:problem}. 
	\item We holistically describe our CAMEO framework for solving the lossy compression problem under constraints of statistical features in Section~\ref{sec:algorithm}. Key aspects include general methodology, incremental maintenance of ACF and PACF, and blocking and parallelization strategies.
	\item We conduct broad experiments to study CAMEO compared to state-of-the-art lossy compressors, different datasets, and different time series analytics in Section~\ref{sec:experiments}.
\end{itemize}

\noindent CAMEO yields improvements in compression ratios of 2x on average (and up to 54x on some datasets) while preserving the same deviation of the ACF. Due to the bounded impact on key statistical features, CAMEO better maintains, and sometimes even improves, the forecasting accuracy. Moreover, CAMEO shows promising results in preserving anomaly detection accuracy.
 
\section{Background}
\label{sec:background}

\begin{figure}[!t]
	\resizebox{0.99\linewidth}{!}{
		\footnotesize
		\begin{forest}
			for tree={
				line width=1pt,
				if={level()<2}{
					my rounded corners,
					draw=linecol,
				}{},
				edge={color=linecol, >={Triangle[]}, ->},
				if level=0{%
					font=\bfseries,
					l sep+=.5cm,
					align=center,
					parent anchor=south,
					tikz={
						\path (!1.child anchor) coordinate (A) -- () coordinate (B) -- (!l.child anchor) coordinate (C) pic [draw, angle radius=10mm, every node/.append style={fill=white}, "based on"] {angle};
					},
				}{%
					if level=1{%
						parent anchor=south west,
						child anchor=north,
						tier=parting ways,
						align=center,
						font=\bfseries,
						for descendants={
							child anchor=west,
							parent anchor=west,
							anchor=west,
							align=left,
							l sep=-0.5cm, 
						},
					}{
						if level=2{
							shape=coordinate,
							no edge,
							grow'=0,
							s sep-=.5cm,
							l sep-=.1cm,
							xshift=15pt,
							calign with current edge,
							for descendants={
								parent anchor=south west,
								l sep-=0.5cm
							},
							for children={
								edge path={
									\noexpand\path[\forestoption{edge}] (!to tier=parting ways.parent anchor) -- ([yshift=-15pt]!to tier=parting ways.parent anchor) |- (.child anchor)\forestoption{edge label};
								},
								font=\bfseries,
								for descendants={
									l=0.5cm, 
									no edge,
								},
							},
						}{},
					},
				}%
			},
			[Lossy Compression
			[{\begin{tabular}{@{}c@{}}Functional\\[-0.5em]Approximation\end{tabular}},l sep=0mm  
			[
			[Linear, l=0mm 
			]
			[Non-linear, l=0mm 
			]
			]
			]
			[{\begin{tabular}{@{}c@{}}Domain\\[-0.5em]Transformation\end{tabular}}, l sep=0mm 
			[
			[Frequency, l=0mm
			]
			[Latent Space, l=0mm
			]
			[Trans. Matrix, l=0mm
			]
			]
			]
			[{\begin{tabular}{@{}c@{}}Value\\[-0.5em]Representation\end{tabular}}, l sep=0mm 
			[
			[Dictionary, l=0mm
			]
			[Symbolic, l=0mm
			]
			[Quantization, l=0mm]
			]
			]
			[{\begin{tabular}{@{}c@{}}Line\\[-0.5em]Simplification\end{tabular}}, l sep=0mm
			[
			[Turning Points,l=0mm, tier=tp]
			[{\begin{tabular}{@{}c@{}}Perceptual\\[-0.5em]Imp. Points \end{tabular}},l=0mm, tier=pip, yshift=0.3cm]
			[{\color{red}{\begin{tabular}{@{}c@{}}Statistical\\[-0.5em]Imp. Points \end{tabular}}},l=0mm, tier=sip, yshift=-0.4cm]
			]
			]
			]
	\end{forest}}
\vspace{-0.4cm}
	\caption{Hierarchy for Lossy Time Series Compression.}
	\label{fig:tax}
	\vspace{-0.15cm}
\end{figure}
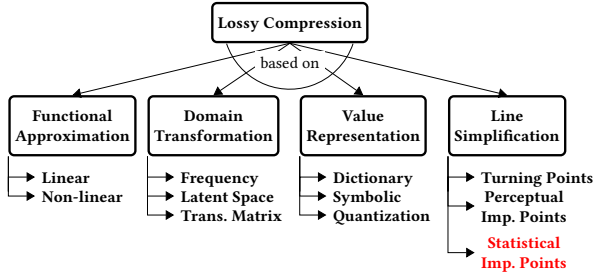

This section describes the necessary background on lossy time series compression (via a hierarchical classification), including the new subcategory \emph{Statistical Important Points}, existing line simplification algorithms, and autocorrelation functions.

\subsection{Lossy Time Series Compression}

Lossy time series compression is the process of converting an input time series $\mat{X}$ of size $n$ into a new compressed representation $\mat{X}^{\prime}$ of size $n^{\prime}$ with $n'\ll n$. The size of $\mat{X}$ and $\mat{X}^{\prime}$ can be measured by the number of bits required to store the time series (e.g., quantization) or the number of elements that need to be stored (e.g., line simplification). The compression ratio $c = \frac{n}{n'}$ measures the effectiveness of compression techniques. Depending on the type of compression, we can reconstruct the original time series with or without loss of information. 
The data distortion measures the loss of information of the reconstructed data compared to the original time series. Finally, we refer the interested reader to existing state-of-the-art surveys~\cite{new_ts_survey} and experimental paper analysis~\cite{DBLP:conf/secon/BoseBKBP16,DBLP:journals/tkde/HungJA13}.

\textbf{Lossy Compression Categories:} There is a plethora of lossy time series compression methods with different trade-offs. Figure~\ref{fig:tax} shows the major categories arranged in a type hierarchy.
\begin{itemize}
	\item \textbf{Functional Approximation:} The data is approximated by one or more functions~\cite{DBLP:conf/sigmod/NgC04,eichinger2015time,keogh2001dimensionality,lazaridis2003capturing,DBLP:conf/vldb/Chen0LLY07,DBLP:conf/icde/BuragohainSS07,DBLP:journals/pvldb/ElmeleegyECAZ09,DBLP:conf/sigmod/KeoghCMP01,KitsiosLPK23}, where the time series is divided into segments, and we store parameters of a low-order polynomial per segment. Such methods guarantee a maximum distortion error per value.
	\item \textbf{Domain Transformation:} The data is transformed into a different mathematical domain~\cite{inamura2003keyframe,DBLP:journals/cacm/BusingerG69,DBLP:journals/tsg/SouzaAP17,DBLP:conf/icpr/LewandowskiRMN10,karim2011wavelet,cooley1965algorithm,ahmed1974discrete}. We then compress the data by retaining significant components in this new domain and discarding less important ones. 
	\item \textbf{Value Representation:} The data is substituted with another more compact representation~\cite{khelifati2019corad,lin2007experiencing,shieh2008sax,marascu2014tristan,gray1998quantization} (e.g., binning or quantization). Here, the compression is achieved by limiting the number of distinct items---and thus, codeword size---while controlling the reconstruction error.
	\item \textbf{Line Simplification:} All points are ranked according to a certain criterion and removed in reverse order~\cite{Yin2011tp,DBLP:journals/eaai/SiY13,DBLP:conf/dexa/SunS20,Hafeez22dpip,fu2017improvement,fu2004specialized,chung2001flexible}. The ranks are updated as points are removed.
\end{itemize}

\begin{figure}[!t]
\centering
	\includegraphics[scale=1]{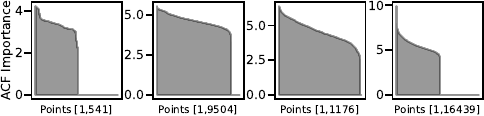}
	\vspace{-0.25cm}
	\caption{ACF Importance Skew (Non-Uniform Importance)}\label{fig:rank}
\end{figure}

\textbf{CAMEO Positioning:} Our method is inspired by existing Line Simplification strategies (which preserve, for instance, the area-under-the-curve). Still, we initiate work on preserving more complex statistical features. We define the category \emph{Statistical Important Points} to group methods that explicitly preserve specific time series statistics such as the ACF and PACF. For instance, Figure~\ref{fig:rank} shows the initial ACF importance skew across points from four time series, highlighting our goal to retain the most important ones. To the best of our knowledge, CAMEO is the first method in this category. 

\subsection{Line Simplification Lossy Compression}
\label{line_simplification}

Lossy methods in this category reduce the number of points while preserving significant characteristics or patterns in the time series. Keeping original data points that are relevant, and discarding the rest (i.e., approximating them via simple imputation techniques), can be advantageous in domains where key data characteristics and indicators need to be preserved. An example application area is material degradation testing in semiconductor manufacturing, where devices are tested with electric stress pulses, and the area-under-the-curve represents the total energy put onto a device.

\textbf{Turning Points:} The central idea behind turning-points (TP) compression is to identify and store only the points at which the time series changes direction, i.e., where it turns from increasing to decreasing, or vice versa~\cite{Yin2011tp,DBLP:journals/eaai/SiY13}. Preserving TPs has the added advantage of more effectively preserving the linear trends of the time series, which is especially relevant for intelligent stock trading~\cite{DBLP:journals/eswa/BaoY08}.

\textbf{Perceptual Important Points:} The core idea behind perceptual-important-points (PIP) compression is to find and store points that are significant or meaningful based on the human perception~\cite{DBLP:journals/eaai/FuCLN08,DBLP:journals/pvldb/JugelJM14}. These points can include peaks, valleys, and other visually salient features that may indicate important events or changes in the data. In detail, a PIP is a point with the maximum distance to an approximation line between two other consecutive PIPs. The set of PIPs is constructed progressively (top-down), starting with an initial approximation line defined by the first and last points. This idea---proposed in 2001~\cite{chung2001flexible}---is very similar to a set of techniques proposed about 30 years earlier for reducing the number of points required for polygonal approximation~\cite{rammer1972iterative,douglas1973algorithms,visvalingam1993line,kronenfeld2020simplification,raposo2013scale}. 

\textbf{VW Algorithm:} We took inspiration from Visvalingam-Whyatt (VW)~\cite{visvalingam1993line} line simplification algorithm to develop CAMEO. VW iteratively removes points based on the areas of the triangles formed by triplets of sequential points. At each iteration, VW replaces the smallest triangle 
with a straight-line segment by eliminating the middle point and computing the areas of the newly formed left and right triangles. The algorithm stops when the smallest triangle's area does not meet the error bound or a specified number of points is removed. This bottom-up approach can be adapted to remove the points with the least impact on any other characteristic to preserve.  

\subsection{Compression Quality Measures}
\label{quality_metrics}

To measure the deviation between the original and reconstructed time series, or the original and reconstructed ACF and PACF, one can use different quality measures $\mathcal{D}(\mat{X},\mat{X}^{\prime})$:
\begin{itemize}
	\item \textbf{Mean Absolute Error:} $\text{MAE} = \frac{1}{n}\sum_{i=1}^{n} |x_i - x^\prime_i|$
	\item \textbf{Root Mean Square Error:} $\text{RMSE} = \frac{1}{n}\sqrt{\sum_{i=1}^{n} (x_i - x^\prime_i)^2}$
	\item \textbf{Normalized RMSE:} $\text{NRMSE} = \frac{\text{RMSE}}{\max(\mat{X})-\min(\mat{X})}$
	\item \textbf{Modified Symmetric Mean Absolute Percentage Error:}
	\[\text{mSMAPE} = \frac{1}{n}\sum_{i=1}^{n}\frac{|x_i - x^\prime_i|}{(|x_i + x^\prime_i|)/2 + S_i}\]
	\noindent where $S_i = \frac{1}{i-1} \sum_{k=1}^{i-1} |x_k - \overline{x}_{i-1}|$ and $\overline{x}_{i-1}= \frac{1}{i-1} \sum_{k=1}^{i-1} x_k$.
\end{itemize}

\subsection{Time Series Autocorrelation Functions}
\label{ts_acf_pacf}

The ACF and PACF are two fundamental statistical concepts that measure the correlation between the observations at a current point in time and observations at different time lags~\cite{box2015time}. 

\textbf{Basic ACF:} The ACF is the Pearson correlation of the time series $\mat{X}$ and a lagged version of itself---computed for lags 1 through a user-provided $L$---and is computed at lag $l$ as follows:
\begin{equation}
	\label{eq:real_acf}
	\textsc{ACF}_l(\mat{X}) = \frac{1}{(n-l)\sigma^2}  \sum\limits_{t=1}^{n-l} (x_{t}-\mu)(x_{t+l}-\mu)  
\end{equation}
where $\mu$ and $\sigma$ are the mean and standard deviation of $\mat{X}$, and $n=\card{\mat{X}}$ (number of points). Equation~\ref{eq:real_acf} assumes the time series is stationary, and thus, $\mu$ and $\sigma$ are the same at all time intervals. If the time series is non-stationary, $\mu$ and $\sigma$ should be computed for $\mat{X}$ and its lagged version $\mat{X}_l$. Specifically, $\mat{X}$ spans $[1 \cdots n-l]$ and $\mat{X}_l$ spans $[l \cdots n]$, thus both time series have $n-l$ elements. 

\textbf{Alternative ACF:} An equivalent formulation of the ACF, but more convenient for later incremental updates, is:
\begin{equation}
	\label{acf_by_expected}
	\textsc{ACF}_l(\mat{X}) = \frac{(n-l)\sum x_t x_{t+l} - \sum x_t \sum x_{t+l}}
	{\sqrt{\left((n-l)\sum x^2_t - (\sum x_t)^2\right)\left((n-l)\sum x^2_{t+l} - (\sum x_{t+l})^2\right)}}  
\end{equation}
whose basic aggregates can be maintained incrementally \cite{DBLP:conf/sigmod/WasayWDI17}.  

\textbf{Basic PACF:} The PACF measures the correlation between current and past observations at lag $l$, removing intermediate lag influences. The \(\textsc{PACF}_l=\phi_{l, l}\), can computed using the Durbin-Levinson (DL)~\cite{durbin,morettin1984levinson} recursion in $\mathcal{O}(L^2)$ as follows:
\begin{equation}
	\label{pacf_dl}
	\phi_{1,1} = \textsc{ACF}_1,\quad 
	\phi_{l,l} = \frac{\textsc{ACF}_l - \sum_{k=1}^{l-1}\phi_{l-1,k}\textsc{ACF}_{l-k}}{1 - \sum_{k=1}^{l-1}\phi_{l-1,k}\textsc{ACF}_k}
\end{equation}
\noindent where $\phi_{l,k} = \phi_{n-1, k} - \phi_{n,n}\phi_{n-1, n-k}$ for $1 \le k \le l-1$.

\textbf{Utility:} The ACF and PACF are valuable tools in time series analytics, often used for understanding the underlying patterns in the series, assisting in selecting the type and order of forecasting models, and enabling precise and reliable forecasts.

\section{Problem Formulation}
\label{sec:problem}

In this section, we introduce three variants for the problem of compressing a time series while preserving statistical features. This problem formulation is independent of concrete algorithms.

\begin{definex}[Statistical Important Points]
	Given a time series $\mat{X}$, an error bound $\epsilon$, a time series statistic $\mathcal{S}$, and a quality measure $\mathcal{D}$, we aim to find a compressed time series $\mat{X}^{\prime}$ (in terms of a subset of original data points) such that:  
	\begin{equation}
		\label{cr_centric_op}
		\begin{aligned}
			\text{max} \quad & \frac{|\mat{X}|}{|\mat{X}^{\prime}|} \\
			\text{s.t.} \quad & \mathcal{D}(\mathcal{S}(\mat{X}), \mathcal{S}(\mat{X}^{\prime})) \leq \epsilon
		\end{aligned}
	\end{equation}
	This optimization objective maximizes the compression ratio between the original time series $\mat{X}$ and its compressed representation $\mat{X}^{\prime}$, while enforcing a bounded deviation of the user-provided statistic $\mathcal{S}$ on the compressed data (measure by $\mathcal{D}$) by at most $\epsilon$.
\end{definex}

\textbf{Complexity:} Similar to other line simplification methods, finding the globally optimal solution efficiently is intractable~\cite{DBLP:conf/esa/KerkhofKLMW19,DBLP:journals/jocg/KreveldLW20}. Thus, we aim to find approximate solutions with suboptimal compression ratios but hard or high-probability guarantees on the deviation from $\mathcal{S}$.
Furthermore, preserving statistical features on window aggregates of the original time series may be more interesting. For example, a time series in 4-second granularity with daily seasonality would require an ACF with \numprint{21600} lags to capture this characteristic. Therefore, optimizing a variant of the \emph{Statistical Important Points} problem might be more meaningful, aiming to preserve statistical features on aggregated time series.

\begin{definex}[Statistical Important Points on Aggregates]
	Given a time series $\mat{X}$, an error bound $\epsilon$, a time series statistic $\mathcal{S}$, a quality measure $\mathcal{D}$, and an additional aggregation function $\textsc{Agg}_\kappa$ over $\kappa$ data points, we aim to find a compressed time series $\mat{X}^{\prime}$ (in terms of a subset of original data points) such that:  
	\begin{equation}
		\label{cr_centric_op_agg}
		\begin{aligned}
			\text{max} \quad & \frac{|\mat{X}|}{|\mat{X}^{\prime}|} \\
			\text{s.t.} \quad & \mathcal{D}(\mathcal{S}(\textsc{Agg}_\kappa(\mat{X})), \mathcal{S}(\textsc{Agg}_\kappa(\mat{X}^{\prime}))) \leq \epsilon
		\end{aligned}
	\end{equation}
	where $\textsc{Agg}_\kappa(\mat{X}) = [a_1, \dots, a_{n/\kappa}]$ and $a_i = \textsc{Agg}_\kappa(x_{[i:i+\kappa]})$. The aggregation function $\textsc{Agg}_\kappa$ needs to be additive, semi-additive, or additively-computable to enable incremental updates~\cite{DBLP:conf/sigmod/WasayWDI17}.
\end{definex}

\begin{example}[SIP on Aggregates] 
To illustrate the SIP on Aggregates problem, assume an original time series $\mat{X}$ sampled every minute, $\epsilon = 0.01$, $\mathcal{S} = \textsc{ACF}$, $\textsc{Agg}_\kappa = \sum_i^{i+30} X_i/30$ (the mean value every $\kappa=30$ minutes), and $\mathcal{D}=\textsc{MAE}$. Here, Equation~\ref{cr_centric_op_agg}'s constraint renders to
\[\textsc{MAE}\left(\textsc{ACF}_{1\cdots L}\left(\frac{\sum_i^{i+30} \mat{X}_i}{30}\right), \textsc{ACF}_{1\cdots L}\left(\frac{\sum_i^{i+30} \mat{X}^{\prime}_i}{30}\right)\right) \leq 0.01,\] 
where $i$ iterates through and aggregates consecutive tumbling (i.e., jumping) windows in the time series. Alternative problem formulations exchange the hard and soft constraints of the above optimization objectives to reach desired compression ratios without unnecessary exploration of parameters.
\end{example}

\begin{definex}[Compression-Centric Statistical Important Points]
	\label{def:data-centric}
	Given a time series $\mat{X}$, a statistic $\mathcal{S}$, a quality measure $\mathcal{D}$, an optional aggregation function $\textsc{Agg}_\kappa$ and a minimum compression ratio $c$, we aim to find a compressed time series $\mat{X}^{\prime}$ such that:
	\begin{equation}
		\label{data_centric_op}
		\begin{aligned}
			\text{min} \quad & \mathcal{D}(\mathcal{S}(\textsc{Agg}_\kappa(\mat{X})), \mathcal{S}(\textsc{Agg}_\kappa(\mat{X}^{\prime}))) \\
			\text{s.t.} \quad & \frac{|\mat{X}|}{|\mat{X}^{\prime}|} \geq c
		\end{aligned}
	\end{equation}
	This optimization objective minimizes the distortion between the original and reconstructed statistical features while removing points until the compression ratio $c$ is reached. 
\end{definex}

\section{\textit{CAMEO} Framework}
\label{sec:algorithm}

\begin{figure}[!t]
	\centering
	\includegraphics[width=0.21\textwidth]{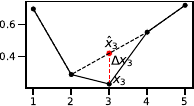}
	\hfill
	\includegraphics[width=0.21\textwidth]{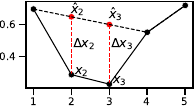}
	\vspace{-0.25cm}
	\caption{Linear Interpolation of x\textsubscript{3} (Left) 
		and x\textsubscript{2} (Right).}\label{fig:interp}
\end{figure}

CAMEO employs a greedy iterative approach to eliminate points from the time series to solve the \emph{Statistical Important Points} problem. At each iteration, the point with the minimum impact on the ACF is removed (throughout this section, we will use only the ACF for simplicity). To compute the impact on the ACF, we assume that the removed point is replaced with a linearly interpolated value. The difference between the interpolated value and the original (denoted as $\Delta x_3$) allows us to recompute the ACF in $\mathcal{O}(L)$ time by updating simple aggregates. Figure~\ref{fig:interp} (left) shows how $x_3$ is removed and replaced with $x^\prime_3$ through interpolation using $x_2$ and $x_4$. After updating the ACF, we check the constraint and continue removing points as long as the error bound is not exceeded.
In this section, we introduce the overall compression algorithm, as well as three techniques for improving its runtime efficiency: incremental maintenance of the ACF, blocking of local neighborhoods, and different parallelization strategies. The underlying greedy heuristics of our algorithm include (1) selecting the next best point and (2) updating the ACF impact in a local neighborhood.

\subsection{Overall Compression Algorithm}

CAMEO starts by computing the ACF of the original time series and its aggregates as shown in Algorithm~\ref{alg:inc_acf_pacf}. We use these aggregates to compute the impact of removing each point on the ACF, assuming its reconstruction by linear interpolation. The interpolation error is used to update the aggregates, allowing the ACF to be recalculated incrementally. This iterative process ensures that computational overhead remains manageable. We provide more details in Section~\ref{sec:inc_acf_pacf}.

\begin{algorithm}[!t]
	\caption{CAMEO}\label{alg:inc_acf_pacf}
	\begin{algorithmic}
		\Require{Time Series $\mat{X}$, Error Bound $\epsilon$, Max Lag $L$}
		\Ensure{List of Remaining Points $\mat{X}'$}
		\State $\textsc{ACFAgg} \gets \textsc{ExtractAggregates}(\mat{X})$ \Comment{Get ACF aggregates}
		\State $\textsc{P}_\textsc{L} \gets \textsc{GetACF}(\textsc{ACFAgg})$ \Comment{Get raw ACF}
		\State $\textsc{H} \gets \textsc{GetAllImpact}(\textsc{ACFAgg}, \mat{X})$ \Comment{Heap of distortions}
		\While{$\textsc{Top}(\textsc{H}) \neq \textsc{NULL}$} \Comment{Not empty}
		\State $\mat{x}_i \gets \textsc{Pop}(\textsc{H})$                \Comment{Get next point}
		\State $\Delta \mat{x}_i \gets \textsc{Interpolate}(\mat{x}_i)$                \Comment{Get interpolation error}
		\State $\textsc{ACFAgg} \gets \textsc{Update}(\textsc{ACFAgg}, \Delta \mat{x}_i)$ \Comment{Update $\textsc{ACFAgg}$}
		\State $\hat{\textsc{P}}_L \gets \textsc{GetACF}(\textsc{ACFAgg})$ \Comment{Get new ACF}
		\If{$\mathcal{D}(\hat{\textsc{P}}_\textsc{L}, \textsc{P}_\textsc{L}) \ge \epsilon$}\label{line:condition} \Comment{Check error bound}
		\State \Return $\mat{X}'$  \Comment{Error bound reached}
		\EndIf
		\State $\mat{X}' \gets \textsc{Remove}(\mat{X}, \mat{x}_i)$ \Comment{Remove the point}
		\State $\textsc{H} \gets \textsc{ReHeap}(\textsc{H}, \mat{x}_i)$ \label{ln:summary:reheap} \Comment{Update impact of points in $\mathcal{N}_h(x_i)$}
		\EndWhile
		\State \Return $\mat{X}'$
	\end{algorithmic}
\end{algorithm}

\textbf{Initialization:} To rank points for removal, we start by computing the impact on the ACF of removing each point. The first and last points cannot be removed; thus, their impact is set to infinite. We call the function \textsc{GetAllImpact}, which is shown in Algorithm~\ref{alg:extract_all}. Given the ACF aggregates, the impact for each point is computed in parallel by a few arithmetic operations and can be vectorized via matrix operations. Finally, the function returns a heap $\textsc{H}$ with the computed values. The time complexity of this function is $\mathcal{O}(L n + n)$. The first term is the computation of the impact on the ACF, and the second is Floyd's method~\cite{floyd1964algorithm} to \emph{heapify} the computed values. 

\textbf{Inner Loop and Updating Heuristic:} Every time a point $\mat{x}_i$ is popped from the heap, we compute the interpolation error ($\Delta \mat{x}_i$), update the ACF aggregates (\textsc{Update}) and recompute its value ($\hat{\textsc{P}}_L$). We continue by checking the constraint and removing the points if the constraint is not exceeded. Removing and interpolating points can alter the relationships between data points at different lags, affecting the ACF and invalidating previously computed impact values. To address this, we recalculate the impacts of neighboring points of $\mat{x}_i$ using the $\textsc{ReHeap}$ operation. This ensures that local updates maintain consistency in the ACF and prevent the propagation of errors to future iterations. This heuristic---which resembles a blocking strategy---balances computational efficiency and accuracy in updating the ACF. In Section~\ref{sec:blocking}, we provide the necessary implementation details for this heuristic.

\begin{algorithm}[!t]
	\caption{Get ACF Impact for All Points}\label{alg:extract_all}
	\begin{algorithmic}
		\Require{ACF aggregates $\textsc{ACFAgg}$, Time Series $\mat{X}$, Raw ACF $P_L$}
		\Ensure{Heap with Impact on ACF per each Point $H$}
		\State $i \gets [1, \cdots, n-1]$ \Comment{Get indices}
		\State $l \gets [1, \cdots, L]$ \Comment{Get lags}
		\State $n \gets [n-1, \cdots, n-L]$ \Comment{Get size for all lags}
		\State $\Delta \mat{X} \gets \frac{1}{2}(\mat{X}[2:]-\mat{X}[:-2])-\mat{X}[i]$ \Comment{Get all deltas $x_i$ by LP}
		\State $sx \gets \textsc{ACFAgg}.sx + \Delta \mat{X}$  \Comment{$\sum x$}
		\State $sx_l \gets \textsc{ACFAgg}.sx_l + \Delta \mat{X}$  \Comment{$\sum x_l$}
		\State $sx^2 \gets \textsc{ACFAgg}.sx^2 + \frac{1}{n}\Delta \mat{X} (\Delta \mat{X} + 2\mat{X}[i])$ \Comment{$\sum x^2$}
		\State $sx^2_l \gets \textsc{ACFAgg}.sx^2_l + \frac{1}{n}\Delta \mat{X} (\Delta \mat{X} + 2\mat{X}[i])$ \Comment{$\sum x^2_l$}
		\State $sxx_l \gets \textsc{ACFAgg}.sxx_l + \frac{1}{n}\Delta \mat{X} (\mat{X}[i - l] + \mat{X}[i + l])$  \Comment{$\sum xx_l$} 
		\State $\hat{\textsc{P}}_{L} \gets \textsc{GetACF}(sx, sx_l,sx^2,sx^2_l,sxx_l)$ \Comment{Apply Equation~\ref{acf_by_expected}}
		\State $\textsc{H} \gets \textsc{Heapify}(\mathcal{D}(\hat{\textsc{P}}_{L}, \textsc{P}_L))$ \Comment{Floyd's method}
		\State \Return $\textsc{H}$
	\end{algorithmic}	
\end{algorithm}

\textbf{Decompression:} Like many line-simplification algorithms, CAMEO employs linear interpolation for decompression. Consequently, the reconstructed time series is a set of linear functions. This process is simple and efficient, requiring only a single forward pass over the remaining points. This assumption allows us to compute the impact on the ACF and PACF incrementally. 

\vspace{-0.3in}
\subsection{Incremental ACF and PACF}
\label{sec:inc_acf_pacf}
Computing the ACF or PACF from scratch for every removed point is infeasible for large time series. Hence, we incrementally maintain the autocorrelation functions---for constraint validation during compression---by keeping track of Equation~\ref{acf_by_expected}'s basic aggregates: 
\begin{equation}
	\vspace{-0.07in}
	\begin{aligned}
		sx        & = \sum_{i=0}^{n-l} x_i               & sx_l      & = \sum_{i=l}^{n} x_i       & sxx_l  & = \sum_{i=0}^{n-l} x_ix_{i+l} \\
		sx^2      & = \sum_{i=0}^{n-l} x_i^2             & sx^2_l    & = \sum_{i=l}^{n} x^2_i     &
	\end{aligned}
	\label{eq:basic_agg}
\end{equation}
These aggregates are computed by the function \textsc{ExtractAggregates} per lag $l\in[1,L]$ in Algorithm \ref{alg:inc_acf_pacf} with complexity $\mathcal{O}(L n)$ dominated by $sxx_l$. Later, when removing the point $x_i$, we compute the distance $\Delta x_i$ between $x_i$ and its interpolated value $\hat{x}_i$, i.e., $\Delta x_i = \hat{x}_i - x_i$. Given $\Delta x_i$, we derive the following update rules:
\begin{equation}
	\begin{split}
		sx & += \Delta x_i, \quad sx^2 += \Delta x_i(2x_i + \Delta x_i), \quad sx_l += \Delta x_i \\
		sx^2_l & += \Delta x_i(2x_i + \Delta x_i), \quad sxx_l += \Delta x_i (x_{i-l} + x_{i+l})
	\end{split}  
	\label{eq:simple_update}      
\end{equation}
Once the aggregates are updated, we can compute the ACF at a specific lag using Equation~\ref{acf_by_expected}. Similarly, to compute and preserve the PACF, we incrementally maintain the ACF and apply the DL recursion in Equation~\ref{pacf_dl} albeit with higher computational cost.

\textbf{Update Rules for Multiple Elements:} In some cases, removing a point requires interpolating more than one element, as shown in Figure~\ref{fig:interp} (right). In that case, the basic aggregates are updated by summing over the deltas of every interpolated point. Specifically, if removing point $x_i$ requires interpolating the $m$ points $[x_j, \ldots, x_i, \ldots, x_{j+m}]$ (changed interpolations until the next remaining points left and right), the update rules are:
\begin{equation}
	\vspace{-0.1in}
	\begin{gathered}
		\begin{aligned}
			sx        & += \sum_{k=j}^{j+m} \Delta x_k,                                      \\
			sx_l   & += \sum_{k=j}^{j+m} \Delta x_k,                                      \\
		\end{aligned}
		\qquad
		\begin{aligned}
			sx^2      & += \sum_{k=j}^{j+m} \Delta x_k(2x_k + \Delta x_k),                   \\
			sx^2_l & += \sum_{k=j}^{j+m} \Delta x_k(2x_k + \Delta x_k),                   \\
		\end{aligned}\\
		sxx_l   += \sum_{k=j}^{j+m} \Delta x_k (x_{k-l} + x_{k+l}) + \sum_{k=j}^{j+m-l} \Delta x_k\Delta x_{k+l}
	\end{gathered}
	\label{eq:update_m}
\end{equation}
Ideally, updating the basic aggregates should not materialize the interpolation of the points from $j$ to $m$ because they are affine functions. 
However, there is no straightforward way to update $sxx_l$ without any assumption on the time series. Note that updating $sxx_l$ has a time complexity of $\mathcal{O}(mL)$ since it calculates a distinct value for each $k$, influenced by $l$. The rest of the basic aggregates can be updated in $\mathcal{O}(L + m)$.

\textbf{Update Rules with Aggregation Function:} Solving the \emph{Statistical Important Points on Aggregates} problem requires additional extensions. Given the aggregation function $\textsc{Agg}_\kappa$, we first compute and store all $a_i \in \textsc{Agg}_\kappa(X)$. Subsequently, while removing the points $x_i$, we incrementally update the aggregates:
\begin{equation}
\vspace{-0.1in}
	\begin{aligned}
		sa        & = \sum_{i=0}^{\lfloor n/\kappa\rfloor-l} a_i               & sa_l      & = \sum_{i=l}^{\lfloor n/\kappa\rfloor} a_i       & saa_l  & = \sum_{i=0}^{\lfloor n/\kappa\rfloor-l} a_i a_{i+l} \\
		sa^2      & = \sum_{i=0}^{\lfloor n/\kappa\rfloor-l} a_i^2             & sa^2_l    & = \sum_{i=l}^{\lfloor n/\kappa\rfloor} a^2_i     &
	\end{aligned}
	\label{eq:basic_agg_agg}
\end{equation}
When removing the point $x_i$, we again consider two cases. First, if only one point is interpolated, the update rules are:
\begin{equation}
\vspace{-0.1in}
	\begin{split}
		sa & += \Delta a_i, \quad sa^2 += \Delta a_i(2a_j + \Delta a_i), \quad sa_l += \Delta a_i \\
		sa^2_l & += \Delta a_i(2a_i + \Delta a_i), \quad saa_l += \Delta a_i (a_{i-l} + a_{i+l})
	\end{split}  
	\label{eq:simple_update_agg}      
\end{equation}
where $\Delta a_i = \textsc{Agg}_\kappa([\hat{x_i}, \ldots, x_{i+\kappa}]) - a_i$. Note, if $\textsc{Agg}_\kappa$ is commutative and associative, it is possible to avoid computing $\textsc{Agg}_\kappa$ overall points avoiding computations. For example, if $\textsc{Agg}_\kappa$ is the mean function, then $\Delta a_i = (\hat{x}-x_i)/\kappa$ or if the function is the maximum, then $\Delta a_i = \max(\hat{x}_i, a_i) - a_i$. Second, if $m$ points are interpolated, we first compute all $\Delta a_i$ by mapping the interpolated points $x_i \in [x_{j}, \ldots, x_{j+m}]$ to the aggregate $a_i$ the point is associated with. Then, $\Delta a_i = \textsc{Agg}_\kappa(\hat{x}_i, \hat{x}_{i+1}, \ldots, x_{i+\kappa}) - a_i$, which requires recomputing $\textsc{Agg}_\kappa$ for all elements, if all elements are interpolated. After computing all $\Delta a_i$, similar to Equation~\ref{eq:simple_update_agg}, we reuse Equation~\ref{eq:update_m} but utilize the aggregates instead of the data points in $\mat{X}$.

\subsection{Blocking} 
\label{sec:blocking}

Blocking is a well-known strategy in entity resolution (deduplication) to enhance efficiency by limiting comparisons to entities within the same bucket. Inspired by this concept, CAMEO updates the ACF impact only for the neighboring points surrounding the removed point rather than updating all points in the time series.

\textbf{Blocking Heuristic:} CAMEO's blocking heuristic relies on the assumption of temporal locality. We assume that removing a point affects the nearby points, and its impact diminishes as we move further away. Thus, we update the impact on the ACF of only $h$-neighboring points of $\mat{x}_i$. Our heuristic ignores computations of distance points to reduce the computational cost without unnecessarily sacrificing the compression ratio. Our experiments show that $h$ can be set to $\log n$, reducing time complexity from $\mathcal{O}(n^2)$ to $\mathcal{O}(n \log n)$. Interesting future work includes other blocking strategies, such as locality based on correlated lags (e.g., seasonality).

\begin{figure}[!t]
	\centering
	\includegraphics{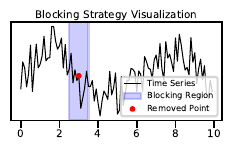}
	\hfill
	\includegraphics{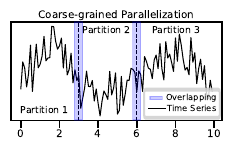}
	\vspace{-0.25cm}
	\caption{Blocking and Coarse-grained Parallelization.}\label{fig:strategies}
\end{figure}

\textbf{Implementation Details:} To identify the $h$-neighboring points of the removed point $\mat{x}_i$, we use the left and right pointers associated with each point in the heap. Starting from the immediate neighbors of $\mat{x}_i$, we traverse up to $h$ hops in both directions, collecting valid (non-removed) neighbors. Whenever a point is removed, the left and right neighbors' pointers are updated to maintain. Figure~\ref{fig:strategies} (Left) illustrates this concept. The \textsc{ReHeap} procedure then assesses their impact on the ACF via the update rules in Equations~\ref{eq:simple_update} and~\ref{eq:update_m}. If the recalculated impact differs from the initial value, an update in the heap is performed in $\mathcal{O}(\log n)$. 

\subsection{Parallelization}

When dealing with very large time series, the application of parallelization strategies becomes indispensable to significantly improve computational time. 
In CAMEO, we implement two different parallelization strategies, namely, \emph{fine-grained} and \emph{coarse-grained} approaches, with their specific advantages and disadvantages.

\textbf{Fine-grained Parallelization:} The first strategy is designed with the primary objective of improving runtime efficiency without additional heuristics that might impact the quality of compression. In CAMEO's blocking strategy, each neighbor's impact on the ACF can be calculated independently during the \emph{look-ahead} phase. Given $T$ threads, we segment the number of neighbors, $h$, into static chunks of size $h/T$, assigning each chunk to a thread. Each thread independently computes the \emph{look-ahead} impact on the ACF for its designated chunk, thereby reducing execution time. However, there are fine-grained synchronization barriers for every removed point.

\textbf{Coarse-grained Parallelization:} This strategy is designed for systems with many cores and very large time series data. The core idea is to partition the time series $\mathbf{X}$ into $T$ consecutive chunks, assign a thread to each partition, and compress each partition independently using the single-threaded CAMEO algorithm. Each partition independently computes and updates its own aggregates while concurrently handling overlapping regions. Synchronization overhead is minimized by allowing each thread to work independently within a local ACF deviation threshold of $p \cdot \epsilon / T$. Once a partition reaches its local error threshold, synchronization begins to update global aggregates across all partitions. This way, we introduce synchronization only when necessary to guarantee the global allowable ACF deviation is not exceeded. 

\begin{example}[Coarse-grained Parallelization] 
Consider a time series $\mathbf{X}$ divided into three partitions $P_1$, $P_2$, and $P_3$, as shown in Figure~\ref{fig:strategies} (Right). Let $sx(P_i)$ denote the sum of all points of partition $P_i$. Each partition computes and updates its own aggregates independently: $sx(P_1)$, $sx(P_2)$, and $sx(P_3)$. Similarly, the aggregates $sx_l(P_i)$, $sx^2(P_i)$, and $sx^2_l(P_i)$ are handled independently within each partition $P_i$. For the dot product between the lagged time series, the overall aggregate $sxx_l$ is computed from the aggregates per partition $sxx_l(P_i)$ and the contributions of overlapping regions: $sxx_l = \sum_{i=1}^3 sxx_l(P_i) + sxx_l(\text{Overlap}_{12}) + sxx_l(\text{Overlap}_{23})$, where $sxx_l(\text{Overlap}_{ij}) = \sum_{t \in P_i, t+l \in P_j} x_t x_{t+l}$ accounts for cross-products where $x_t$ is in $P_i$ and $x_{t+l}$ is in $P_j$. Only the threads handling $P_i$ and $P_{i+1}$ need to synchronize when accessing the small overlapping regions, and thus, synchronization overhead is negligible. Given these aggregates, each partition can operate independently until meeting the error bound $p\cdot \epsilon/3$. Once a partition reaches its local error threshold, the global ACF can be computed by synchronizing access to the aggregates across all the partitions, ensuring that the overall ACF deviation remains within the specified error bound $\epsilon$. 
\end{example}

\begingroup
\renewcommand{\arraystretch}{0.99} 
\begin{table*}[t]
	\centering
	\small \setlength\tabcolsep{6.5pt}
	\caption{Datasets Summary. The standard deviation is denoted by $\sigma$, and the probability of a data point to be higher than, equal to, or lower than that of the previous data point is denoted by ($\mathbf{p_\downarrow}$), ($\mathbf{p_=}$) and ($\mathbf{p_\downarrow}$).}
	\vspace{-0.4cm}
	\label{tab:description}
	\begin{tabular}{cccccccccccc} 
		\toprule
		\textbf{\textbf{Dataset}} & \textbf{Length} & \begin{tabular}[c]{@{}c@{}}\textbf{ACF}\\\textbf{\#Lag}\end{tabular} & \textbf{ACF1} & \textbf{ACF10} & \textbf{PACF5} & \begin{tabular}[c]{@{}c@{}}\textbf{Min}\\\textbf{Value}\end{tabular} & \multicolumn{1}{l}{\textbf{Range}} & \textbf{Median} & $\mathbf{\sigma}$ & \textbf{$\mathbf{p_\uparrow}\ $---$\ \mathbf{p_=}\ $---$\ \mathbf{p_\downarrow}$} & \begin{tabular}[c]{@{}c@{}}\textbf{Mean }\\\textbf{Delta}\end{tabular}  \\ 
		\midrule
		ElecPower~\cite{hepc}                 & 2,977 & 48                                                                   & 0.768         & 3.38           & 0.94           & 0.099                                                                & 5.7                                & 0.29            & 0.74              & 48\%–0\%–52\%                                                         & 8e-04                                                                   \\
		MinTemp~\cite{melbourne-temps}                   & 3,652           & 365                                                                  & 0.774         & 5.97           & 1.32           & 0.01                                                                 & 26.3                               & 11.0            & 4.01              & 52\%–1\%–47\%                                                         & 0.002                                                                   \\
		Pedestrian~\cite{DBLP:conf/nips/GodahewaBWHM21}                & 8,766           & 24                                                                   & 0.896         & 1.02           & -0.11          & 0.00                                                                 & 5,573                              & 396             & 1,017             & 45\%–0\%–55\%                                                         & 0.004                                                                   \\
		UKElecDem~\cite{ukelecdem}                 & 17,520          & 48                                                                   & 0.988         & 7.2            & 0.37           & 16,309                                                               & 30,124                             & 27,857          & 6,071             & 44\%–0\%–56\%                                                         & 0.34                                                                    \\ 
		\midrule
		AUSElecDem~\cite{DBLP:conf/nips/GodahewaBWHM21}                & 230,736         & 7 on 48                                                              & 0.762         & 5.09           & 1.09           & 3,498                                                                & 9,367                              & 6,783           & 1,361             & 42\%–0\%–58\%                                                         & 0.001                                                                   \\
		Humidity~\cite{humidity-data}                  & 397,440         & 24 on 60                                                             & 0.951         & 2.66           & -0.07          & 12.65                                                                & 87.27                              & 76.38           & 19.73             & 55\%–3\%–42\%                                                         & 5e-06                                                                   \\
		IRBioTemp~\cite{temp-bio-data}                 & 878,400         & 24 on 60                                                             & 0.958         & 4.41           & 0.17           & -5.47                                                                & 54.6                               & 23.21           & 8.55              & 45\%–5\%–50\%                                                         & -3e-06                                                                  \\
		SolarPower~\cite{DBLP:conf/nips/GodahewaBWHM21}                & 986,297       & 24 on 120                                                            & 0.90         & 1.02           & 0.125          & 0.00                                                                 & 116.5                              & 0.0             & 43.33             & 12.5\%–75\%–12.5\%                                                          & 0.0                                                                    \\
		\bottomrule
	\end{tabular}
	\vspace{-0.25cm}
\end{table*}
\endgroup

\section{Experiments}
\label{sec:experiments}

Our experiments study CAMEO's compression ratios, incurred errors, and runtime characteristics on various datasets and models and compare them with lossless and lossy compression techniques.

\subsection{Experimental Setup} We conduct the experiments on a Linux server running an Intel(R) Xeon(R) Gold 128-core processor at 0.8\,GHz with 120\,MB L3 cache and 1\,TB of main memory. We implemented CAMEO in Cython 3.0.0, compiled with GCC version 9.4.0 at optimization level O3 and OpenMP version 4.5. Cython provides considerable performance enhancements through static typing of variables while supporting NumPy~\cite{harris2020array}. The Cython framework further avoids the problems of Python's Global Interpreter Lock (GIL) and takes full advantage of multithreading. By default, our experiments use the \emph{Statistical Important Points on Aggregates} problem (Equation~\ref{cr_centric_op_agg}), the mean as aggregation function $\textsc{Agg}_\kappa$, and quality measure $\mathcal{D}=\mathrm{MAE}$. 

\textbf{Line-Simplification Baselines:} We assess CAMEO's performance against three line-simplification algorithms (see Section~\ref{line_simplification}), which we adapted to support the constraint on the ACF: 
\begin{itemize}
	\item the Visvalingam-Whyatt (VW) algorithm~\cite{visvalingam1993line}, 
	\item the Turning Points (TP) algorithm~\cite{DBLP:journals/eaai/SiY13} (with two evaluation functions: Sum of the Absolute Values (TPs), and Mean Absolute Error (TPm)~\cite{DBLP:journals/eaai/SiY13}), and 
	\item the Perceptual Important Points (PIP) algorithm~\cite{DBLP:journals/eaai/FuCLN08,DBLP:journals/pvldb/JugelJM14} (with two importance functions: Vertical (PIPv) and Euclidean (PIPe) distances~\cite{DBLP:journals/eaai/FuCLN08}).
\end{itemize}

\textbf{Additional Baselines:} We also compare CAMEO with three well-known lossy compression algorithms: Poor Man's Compression Mean (PMC)~\cite{lazaridis2003capturing}, Swing Filter (SWING)~\cite{DBLP:journals/pvldb/ElmeleegyECAZ09}, Sim-Piece (SP)~\cite{KitsiosLPK23}, and the Fast Fourier Transform (FFT)~\cite{cooley1965algorithm}. PMC, SWING, and SP learn constant and linear approximations which are prevalent Functional Approximation approaches. FFT can compress the data by discarding the less important high-frequency components of the frequency spectrum. Since enforcing the ACF constraint while compressing is not straightforward, we perform a trial-and-error exploration of the parameters of these methods while recording the ACF deviation. Finally, we also compare with Gorilla~\cite{pelkonen2015gorilla} and Chimp~\cite{liakos2022chimp}, as lossless compression. Both use XOR to reduce the number of bits needed to represent the data, assuming that consecutive points are likely to be similar or equal. We use the bits per value $Bits/v=Bits(\mat{X}^{\prime})/|\mat{X}|$ metric to measure the effectiveness of the compression---where $Bits(\mat{X}^{\prime})$ denotes the number of bits required to store the compressed representation. 

\textbf{Datasets:} We use eight publicly available datasets. Our primary selection criterion was the presence of a seasonal component, which is discernible via the ACF. This seasonal component was then used to guide dataset-specific configurations of the number of lags. Table~\ref{tab:description} summarizes the main characteristics of all datasets: 

\begin{itemize}
	\item \textbf{ElecDem}~\cite{hepc}: contains measurements of electric power consumption in one household with a 15-minute sampling rate from 07-2007.
	\item \textbf{MinTemp}~\cite{melbourne-temps}: contains daily min temperature in Melbourne (Australia) from 1981 through 1990.
	\item \textbf{Pedestrian}~\cite{DBLP:conf/nips/GodahewaBWHM21}: contains hourly pedestrian counts of 66 sensors in Melbourne from May 2009 through 05-2020. 
	\item \textbf{UKElecDem}~\cite{ukelecdem}: contains the national electricity demand every half-hour of Great Britain for 2021.
	\item \textbf{AUSElecDem}~\cite{DBLP:conf/nips/GodahewaBWHM21} contains the electricity demand every half-hour in Victoria (Australia) from 2002 to 2015.
	\item \textbf{Humidity}~\cite{humidity-data}: contains relative humidity measurements averaged over 1 minute from 04-2015 through 06-2023.
	\item \textbf{IRBioTemp}~\cite{temp-bio-data}: contains biological surface temperature (degrees Celsius) as measured by infrared sensors averaged over 1 minute from 04-2015 through 06-2023.
	\item \textbf{SolarPower}~\cite{DBLP:conf/nips/GodahewaBWHM21}: contains solar power production recorded every 30 seconds from 08-2019 through 06-2020.  
\end{itemize}
We divide the datasets into two groups of four datasets, where for group 1, we preserve the ACF directly; for group 2, we preserve the ACF on window aggregates. Table~\ref{tab:description} also specifies the number of points per window and the number of lags, e.g., 7 on 48 (under the ACF \#Lag column), which means that we aggregate 48 points per window and keep 7 lags in the ACF.

\textbf{Setting the Number of Lags:} For all experiments, we select the number of lags for the ACF based on a full seasonality period of the data. 
For example, MinTemp contains daily temperature measurements with clear yearly seasonality, and thus, we compute the ACF for 365 lags. 
Likewise, we select the aggregation period based on the seasonality we are interested in preserving. For example, Humidity is a dataset with one-minute sample granularity. Thus, we aggregated over an hour and computed the ACF of 24 lags (seasonal cycle of a day) instead of an ACF with \numprint{1440} lags.

\subsection{Compression Ratio}
\label{sec:cr}

\begin{figure}[!t]
	\includegraphics[width=0.99\columnwidth]{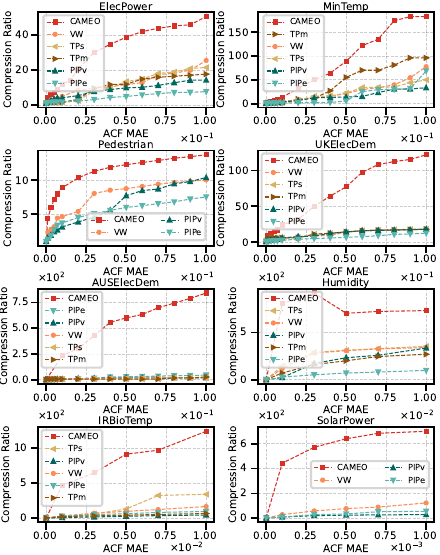}
	\vspace{-0.35cm}
	\caption{Compression Ratio as the ACF Error Increases for Line Simplification Baselines (VW, TP, PIP).}
	\label{fig:acf_cr}
	\vspace{-0.1cm}
\end{figure}

In the first set of experiments, we assess the compression ratio achieved for the eight datasets. Figure~\ref{fig:acf_cr} shows the results CAMEO achieved compared to other line-simplification baselines under varying error bounds for the ACF deviation. CAMEO consistently delivers the best compression ratio among all baselines, mainly because it is the only technique that directly optimizes for the ACF. CAMEO achieves a 1.1x to 54x higher compression ratio, even at very small error bounds. Our extensions of the PIPs and TPs line-simplification methods to constraint the ACF deviation were effective in most datasets except Pedestrian and SolarPower. In these instances, the initial phase of the TP method---which involves removing all non-turning points---results in an ACF that deviates more than the allowed error bounds. Among the line-simplification baselines, VW shows the best performance on average. 

\begin{figure}[!t]
	\includegraphics[width=0.99\columnwidth]{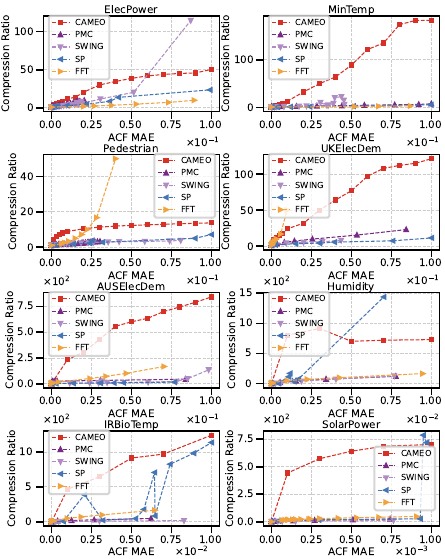}
	\vspace{-0.1in}
	\caption{Compression Ratio as the ACF Error Increases for Lossy Compressor Baselines (PMC, SWING, SP, FFT).}
	\label{fig:lossy_cr}
	\vspace{-0.2cm}
\end{figure}

\textbf{Additional Baselines Results:} We compare CAMEO with the additional lossy compression baselines. As shown in Figure~\ref{fig:lossy_cr}, CAMEO delivers the best compression ratio among all baselines. Some lossy compression methods outperform CAMEO and the remaining baselines in a few instances. For example, FFT outperforms CAMEO in Pedestrian and UKElecDem, which suggests that these datasets predominantly consist of low-frequency components. Similarly, SWING and SP outperform CAMEO on ElecPower and Humidity, showing higher compression ratios at larger error bounds after an initially weaker performance. This higher compression ratio suggests that, in these cases, the error bound is large enough to allow the fitting of a few linear functions without significantly affecting the ACF. Despite these exceptions, CAMEO consistently demonstrates superior compression ratios overall. Notably, SP is not always superior to SWING and PMC when considering their impact on the ACF instead of the error bound. 

\begingroup
\renewcommand{\arraystretch}{0.99} 
\begin{table}[!t]
	\small \setlength\tabcolsep{5.8pt}
	\centering
	\caption{Bits/value Result of Lossless Compression}
	\vspace{-0.35cm}
	\label{tab:bits_value}
	\begin{tabular}{lcccccc} 
		\toprule
		\multicolumn{1}{c}{\multirow{2}{*}{\textbf{Dataset}}} & \textbf{Gorilla} & \textbf{Chimp} & \multicolumn{2}{c}{\textbf{VW}} & \multicolumn{2}{c}{\textbf{Cameo}}  \\ 
		\cmidrule[\heavyrulewidth]{2-7}
		\multicolumn{1}{c}{} & Bits/v    & Bits/v  & Bits/v & $\epsilon$ & Bits/v & $\epsilon$   \\ 
		\toprule
		ElecPower & 63.73     & 55.52   & 23.88  & 3e-3     & 16.03  & 1e-3  \\
		MinTemp     & 59.76     & 22.52   & 16.56  & 7e-3     & 16.31  & 3e-3  \\
		Pedestrian  & 16.63     & 27.45   & 14.45  & 7e-3     & 14.52  & 1e-3  \\
		UKElecDem   & 18.33     & 28.95   & 17.76  & 5e-3     & 7.4    & 1e-3  \\ 
		\midrule
		AUSElecDem  & 56.56     & 53.52   & 49.92  & 1e-4     & 26.63 & 1e-4 \\
		Humidity    & 52.64     & 22.16    & 21.6 & 2e-5 & 1.51 & 1e-5  \\
		IRBioTemp   & 52.40     & 20.33   & 15.13   & 5e-5    & 11.4   & 5e-5     \\
		Solar  & 3.55 & 9.87   & 2.31   & 1e-4 & 0.14	 & 1e-4  \\
		\toprule
	\end{tabular}
\end{table}
\endgroup

\textbf{Bits-per-Value Analysis:} We conclude this analysis by comparing CAMEO and the lossless methods in terms of Bits-per-Value. In detail, we compressed with Gorilla and Chimp, assuming the time series are represented with double precision floating points~\cite{liakos2022chimp} and compute the obtained bits per value metric ($Bits/v$). We compute this metric for CAMEO and VW by multiplying the number of remaining points after compression times 64, which is the number of bits required to store the remaining points. We also report the error bound $\epsilon$ that ensures obtaining a lower $Bits/v$ than both Gorilla and Chimp. As shown in Table~\ref{tab:bits_value}, in all datasets, VW and CAMEO obtain lower $Bits/v$ at reasonably low $\epsilon$, indicating an efficient compression with a very small error on the ACF. Moreover, CAMEO achieves the best compression at smaller $\epsilon$ compared to VW. Overall, Gorilla and Chimp compression ratios are modest, achieving only for half the datasets $Bits/v < 25$ (a compression ratio of $\approx 2.5$). These results show that CAMEO can offer a superior compression ratio while incurring only a small ACF deviation.

\begin{figure}[!t]
	\includegraphics[width=0.99\columnwidth]{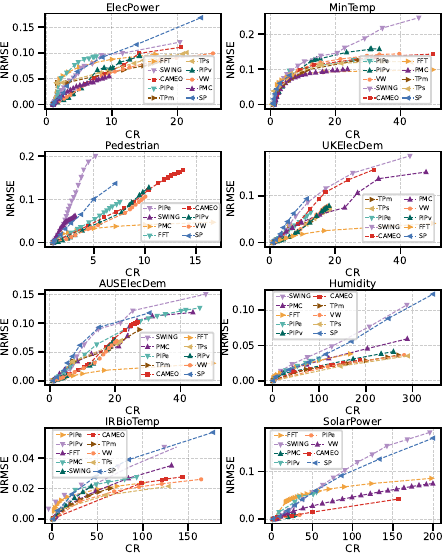}
	\vspace{-0.1in}
	\caption{NRMSE as the Compression Ratio Increases.}
	\label{fig:lossy_cr_nrmse}
\end{figure}

\vspace{-0.1in}
\subsection{Decompression Error}
\label{sec:rq}

In a second set of experiments, we investigate the reconstruction error of the time series after decompression. We compare CAMEO with all baseline methods and report the NRMSE to measure this error. Figure~\ref{fig:lossy_cr_nrmse} illustrates the results. Overall, no single method consistently outperforms the others, with performance highly dependent on the time series characteristics. On average, CAMEO performs on par with the baselines, neither introducing more nor less error for comparable compression ratios. Notably, it never performs the worst and achieves the best results on SolarPower. While CAMEO's average NRMSE is similar to the baselines, it better preserves the ACF.
In contrast, PIPe exhibits the highest NRMSE across multiple datasets among the line-simplification methods. This result indicates that the used Euclidean distance might be an inadequate importance function. When examining the lossy compression techniques, SWING and SP have the highest negative impact on the NRMSE to achieve equivalent compression levels. Notably, FFT performs well in some datasets, highlighting its effectiveness for time series dominated by low-frequency components.

\begingroup
\renewcommand{\arraystretch}{0.99} 
\begin{table*}[t]
	\centering
	\small \setlength{\tabcolsep}{7.0pt} 
	\caption{Compression Times (sec) of the Baselines and Singled-threaded CAMEO (with different blocking sizes).}
	\vspace{-0.4cm}
	\label{tab:exec_time}
	\begin{tabular}{lccccccc|ccccccc} 
		\toprule
		\multicolumn{1}{c}{\multirow{2}{*}{\textbf{Dataset}}} & 
		\multicolumn{1}{c}{\multirow{2}{*}{\textbf{PMC}}} & 
		\multicolumn{1}{c}{\multirow{2}{*}{\textbf{SWING}}} &
		\multicolumn{1}{c}{\multirow{2}{*}{\textbf{SP}}} & 
		\multicolumn{1}{c}{\multirow{2}{*}{\textbf{FFT}}} & 
		\multicolumn{1}{c}{\multirow{2}{*}{\textbf{TP}}} & 
		\multicolumn{1}{c}{\multirow{2}{*}{\textbf{PIP}}} & 
		\multirow{2}{*}{\textbf{VW}} & 
		\multicolumn{7}{c}{\textbf{CAMEO}}  \\ 
		\cline{9-15}
		\multicolumn{1}{c}{}    & \multicolumn{1}{l}{}  & \multicolumn{1}{l}{}  & \multicolumn{1}{l}{}  & \multicolumn{1}{l}{}& \multicolumn{1}{l}{} & \multicolumn{1}{l}{}  &  & $1$ & $\log n$ & $3 \log n$ & $5 \log n$ & $7 \log n$ & $10 \log n$  & $w/b$ \\ 
		\midrule
		ElecPower 	& 6e-4 		& 6e-4 & 2e-3 	& 4e-4 	& 3e-3 & 0.04 	& 0.01 	& 0.02 & 0.02   & 0.09   & 0.2   & 0.33 	& 0.4  & 2.5\\
		MinTemp     & 7e-4    	& 1e-3 & 2e-3	& 4e-4 	& 0.01  & 0.2 	& 0.02 	& 0.06 & 0.07   & 0.6	 & 0.42 	& 1.49 	& 2.75  & 17.7\\
		Pedestrian  & 1e-3    	& 2e-3 & 5e-3	& 1e-3 	& - 	& 0.02  & 8e-3 & 0.03  &  0.09  & 0.2 	 & 0.3 	& 0.5 	& 0.68 &  9.2\\
		UKElecDem   & 3e-3    	& 5e-3 & 1e-2 	& 1e-3  & 8e-3 & 0.05  & 0.02 	& 0.05 & 0.2    & 0.76 	 & 1.08 	& 1.49 	& 2.39  & 64.1\\ 
		\midrule
		AUSElecDem  & 0.04    	& 0.06 & 0.13 	& 0.03    & 0.04    & 0.3   & 0.27 & 0.25  & 0.6 	& 0.3 	& 3.6 	& 8.0 	& 12.2 & \numprint{2554} \\
		Humidity    & 0.04     	& 0.07 & 0.15	& 0.05    & 0.3    & 0.9    & 0.52 	& 1.24 & 10.3  & 21.9 	& 25.3	& 35.6 	& 48.6  & \numprint{6837}\\
		IRBioTemp   & 0.10    	& 0.15 & 0.29 	& 0.16    & 0.25   & 2.1    & 1.6  	& 2.5  & 24.9  & 51.8  & 78.9  & 91.7  & 121    & \numprint{17602}\\
		SolarPower  & 0.11    	& 0.17 & 0.34 	& 0.42    & - 	   & 21   	& 0.72 	& 1.19 &13.8 	& 32.8  & 44.7 	& 52.7 	& 63.3   &  \numprint{5718}\\
		\bottomrule
	\end{tabular}
	\vspace{-0.2cm}
\end{table*}
\endgroup

\begin{figure}[t]
	\includegraphics[width=0.99\columnwidth]{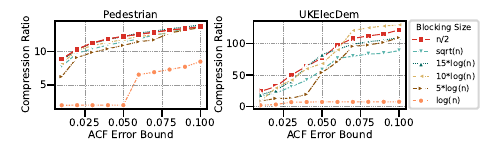}
	\includegraphics[width=0.99\columnwidth]{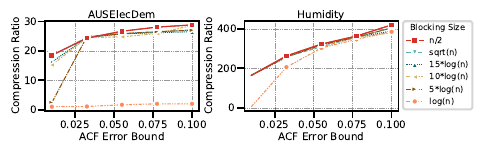}
	\vspace{-0.25cm}
	\caption{Compression Ratio using Blocking.}
	\label{fig:blocking_results}
	\vspace{-0.1cm}
\end{figure}

\subsection{Blocking Strategy}

In a third series of experiments, we conduct micro-benchmarks for the blocking strategy and how it impacts CAMEO's compression performance. We showcase the results using four datasets: Pedestrian, UKElecDem, AUSElecDem, and Humidity (two of both groups). For the case of AUSElecDem and Humidity, the number of blocking hops is multiplied by the size of the aggregation window to cover the necessary lags. Figure~\ref{fig:blocking_results} shows CAMEO's compression ratio as the error bound increases, using different numbers of hops in our blocking strategy. The results show only a slightly reduced compression ratio when using different factors of $\log n$ (from 15 to 5), compared to brute-force updating of many or all neighbors such as $n/2$. In contrast, using $\log n$ results in an inferior compression ratio on all datasets. This observation aligns very well with our previous hypothesis of temporal locality regarding how removing points affects the ACF at different numbers of lags. Thus, considering the size of these datasets, using only $\log n$ does not cover all the lags to properly update the points' impact on the ACF. The following section also shows how the blocking configuration carries through to improved compression time.

\vspace{-0.1in}
\subsection{Compression Time} 
\label{sec:ct}

In a fourth set of experiments, we compare CAMEO's single-threaded compression time with all other baselines. We show the results for an error bound of 0.01 for all the small datasets and 0.001 for the rest. We also enforce a constraint that halts the algorithms once the compression ratio reaches 10. We utilize open-source implementations whenever possible. Specifically, PMC, SWING, SP (TerseTS~\cite{TerseTS}), FFT (NumPy~\cite{DBLP:journals/nature/HarrisMWGVCWTBS20}), and PIP, TP, VW, and CAMEO (implemented by us), all ultimately run as native code. Table~\ref{tab:exec_time} shows the runtime for all baselines and CAMEO as we increase the blocking hops from $1$ to $10\log n$, and without blocking (\textit{w/b}). 

\textbf{Runtime Analysis:} CAMEO's single-threaded implementation performs comparably to other line simplification baselines when using a single hop for blocking. As the number of hops increases, the execution time increases almost linearly. While its execution time increases with higher hops, this trade-off enables CAMEO to achieve significantly higher compression ratios. Removing blocking (\textit{w/b}) makes CAMEO infeasible for real-life applications. Among the baselines, PMC and FFT are the fastest, which is expected considering PMC linear time complexity and FFT's highly optimized implementation. However, they still have the limitation of requiring \textit{trial-and-error} exploration for preserving bounds on the ACF. Finally, TP's initial phase, which preserves only the turning points, positively impacts its execution time, albeit risking not meeting the error-bound guarantee on the ACF. Overall, CAMEO is the preferred choice when optimizing compression ratio over speed while providing strong guarantees on the ACF deviation.

\textbf{PACF Preservation Runtime Analysis:} We also test CAMEO when preserving the PACF. The results show that while the compression ratio is still superior to the baselines, preserving the PACF has a much higher impact on the execution time. For example, when running CAMEO on ElecPower, with blocking at $10\log n$, we obtain an execution time of 2.6 seconds, around 6x slower than preserving the ACF. This increased execution time is due to the quadratic execution time of DL recursion with complexity $\mathcal{O}(L^2)$, computed multiple times per iteration. In future work, we will focus on preserving specific lags to enhance execution speed without sacrificing forecasting accuracy.

\begingroup
\renewcommand{\arraystretch}{0.99} 
\begin{table}
	\centering
	\setlength{\tabcolsep}{6.0pt} 
	\caption{Decompression Times (ms).}
	\vspace{-0.15in}
	\label{tab:dec}
	\begin{tabular}{lccccc} 
		\toprule
		\textbf{Dataset} & \textbf{PMC} & \textbf{SWING} & \textbf{SP} & \textbf{FFT} & \textbf{CAMEO} \\\hline 
		AUSElecDem   & 19 & 17 & 14 & 20  & 12 \\
		Humidity     & 26 & 23 & 18 & 33 &  21    \\
		IRBioTemp    & 80 & 77 & 60 & 97 &  54       \\
		SolarPower   & 88 & 79 & 66 & 352 &  55     \\
		\bottomrule
	\end{tabular}
\end{table}
\endgroup

\begin{figure*}[!t]
	\subfigure[Fine-grained]{
		\label{fig:fine_results}\includegraphics[width=0.91\columnwidth]{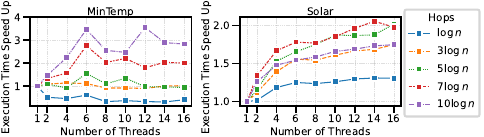}}
	\hfill
	\subfigure[Coarse-grained]{
		\label{fig:coarse_results}
		\includegraphics[scale=0.9]{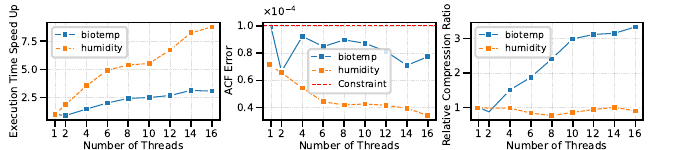}}
\vspace{-0.55cm}	
\caption{Results with Different Parallelization Strategies.}
\vspace{-0.2cm}	
\end{figure*}

\vspace{-0.1in}
\subsection{Decompression Time} 
\label{sec:dec}

In a fifth series of experiments, we evaluate decompression time. Specifically, we compute the execution time of the linear interpolation used as the decompression strategy for CAMEO while directly measuring the decompression times for the other lossy compressors after achieving a 10x compression ratio. Table~\ref{tab:dec} presents the results in milliseconds, with CAMEO representing all line simplification methods due to their identical execution times. The results show that CAMEO achieves significantly faster decompression than the baselines, which is particularly important in scenarios where quick decompression is critical. Interestingly, FFT has the slowest decompression time, contrasting its fast compression performance. This discrepancy arises because the decompression logic of other baselines is much simpler, while FFT retains a complexity of $\mathcal{O}(n \log n)$.

\vspace{-0.08in}
\subsection{Parallelization Strategy}

In a sixth series of experiments, we analyze our coarse- and fine-grained parallelization strategies and end-to-end runtime. 

\textbf{Fine-grained Parallelization Results:} We conduct experiments on MinTemp and Solar. MinTemp exhibits a higher periodic cycle, while Solar is the largest dataset. Figure~\ref{fig:fine_results} shows CAMEO's relative execution time while increasing the number of threads. The parallelization shows moderate runtime improvements across different hop sizes. As threads increase from 1 to 16, the runtime for $\log n$ remains low, particularly for MinTemp, where performance worsens. This result is expected, as the number of neighboring nodes is small, and the overhead of managing multiple threads outweighs the benefits. As the hop size increases, we observe a significant speedup close to 4x for MinTemp at $10\log n$ using 6 threads and a 2x speedup for Solar using 14 threads. An interesting observation is that MinTemp, despite having fewer data points than \textit{Solar}, experiences a greater speedup. This discrepancy is due to MinTemp's higher number of lags (365) compared to Solar's 24. The larger number of lags results in a heavier load for each thread, making parallelization more effective.
On the other hand, Solar's larger size does not translate into significant speedups because the difference in the $\log n$ factor between the two datasets is not large enough to yield substantial improvements. 

\textbf{Coarse-Grained Parallelization:} To showcase the benefits of coarse-grained parallelization, we use the Humidity and IRBioTemp datasets. For both, we set the ACF error bound to 1e-4 and recorded the compression ratio and impact on the overall ACF after merging the results. Figure~\ref{fig:coarse_results} shows the results for increasing numbers of threads. The compression ratio is shown relative to single-threaded execution. 
In both datasets, execution time decreases while maintaining the global ACF deviation within the specified bound. On Humidity, we observe a significant reduction in execution time---up to an 8x speedup---with little change in CR as the number of threads increases. On IRBioTemp, although the execution time improvement is more moderate at 2.5x, we observe a 3x more CR when using 16 threads. The higher CR can be associated with specific characteristics of IRBioTemp that allow local threads to obtain higher CRs before reaching their allowable ACF deviation. Consequently, the overall CR is already higher when synchronization occurs across threads. In conclusion, our coarse-grained parallelization strategy allows users to apply CAMEO to big series.

\begin{figure}[t]
\includegraphics[width=0.9\linewidth,  height=4.69cm]{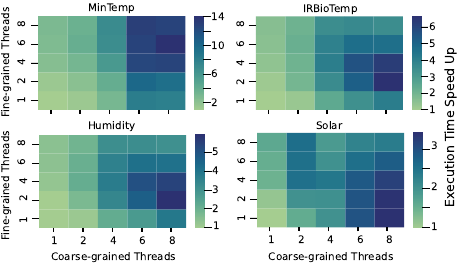}
\vspace{-0.1in}
\caption{Joint Fine- and Coarse-Grained Parallelization.}
\label{fig:all_together_exec_time}
\vspace{-0.1cm}
\end{figure}

\textbf{Hybrid Parallelization:} Figure~\ref{fig:all_together_exec_time} shows the speedup achieved by combining fine- and coarse-grained parallelization strategies in CAMEO. The results are shown for the MinTemp, IRBioTemp, Humidity, and Solar datasets, using a hop size of $10\log n$, corresponding to the longest execution time.
MinTemp shows the most significant improvement among all datasets, with a speedup of up to 14x when using 6 fine-grained and 8 coarse-grained threads. This speedup represents a reduction in execution time from 2.75 seconds (as shown in Table~\ref{tab:exec_time}) to 0.2 seconds, making CAMEO potentially viable for scenarios requiring relatively low-latency compression. While some baseline methods may still be faster, this result highlights the effectiveness of our proposed parallelization strategies.
We observe more moderate improvements for the other datasets, with most of the speedup driven by the coarse-grained strategy. This result aligns with the earlier results in Figure~\ref{fig:fine_results}, as these datasets only need to preserve an ACF of 24 lags. To investigate this further, we conducted an experiment in which we increased the number of lags from 24 to 168 (a week of data) and reran CAMEO with 8 threads for both strategies. The results showed notable improvements: 15x for IRBioTemp, 13x for Humidity, and 7x for Solar. These findings support our hypothesis that the fine-grained strategy becomes more effective as the number of lags increases.
Overall, these results provide a holistic view of the effectiveness of our parallelization strategies in speeding up CAMEO's execution time.

\vspace{-0.1in}
\subsection{Impact on Time Series Forecasting}
\label{sec:forecasting}

In this section, we investigate our original hypothesis---preserving the ACF during compression is beneficial for forecasting analytics. To test this hypothesis, we conduct three experiments. In the first set of experiments (\textbf{EXP1}), we evaluate how different distance metrics $\mathcal{D}$ impact forecasting accuracy using CAMEO and compare the results against the line-simplification baselines. In the second set of experiments (\textbf{EXP2}), we evaluate CAMEO against the rest of lossy compression baselines and three different forecasting models. In the third set of experiments (\textbf{EXP3}), we evaluate CAMEO against our closest line-simplification baseline on highly seasonal time series.

\begin{figure}[!t]
\centering
\subfigure[Forecasting Accuracy of CAMEO and Line-Simplification Baselines.]{\vspace{-0.4in}
	\label{fig:distant_metrics}\includegraphics[scale=0.9]{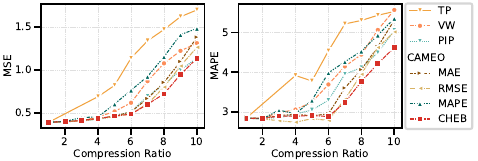}\vspace{-0.4in}}
\subfigure[Forecasting Accuracy of CAMEO and Lossy Compression Baselines.]{\vspace{-0.4in}
	\label{fig:forecasting_impact}\includegraphics[scale=0.33]{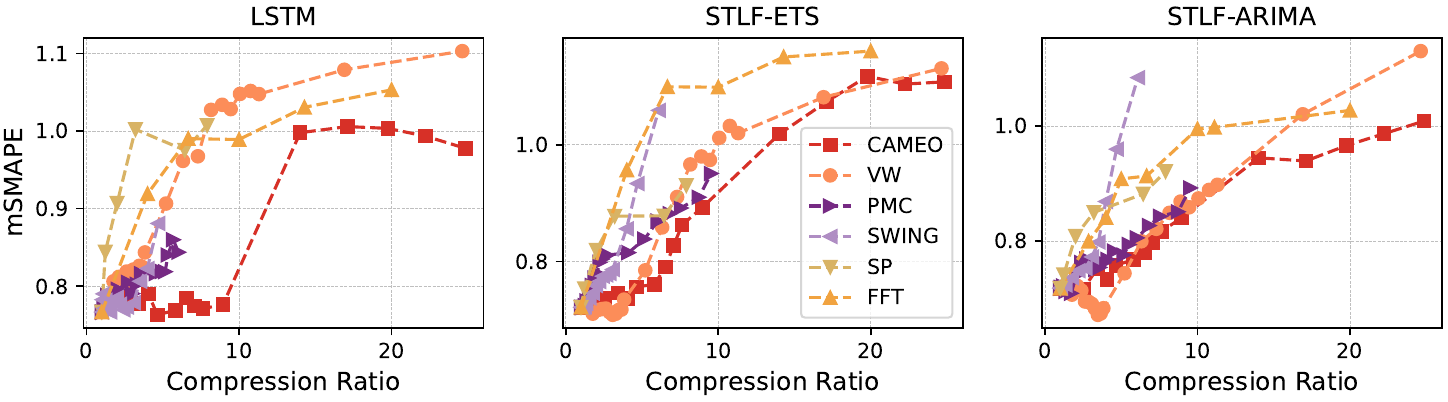}\vspace{-0.4in}}
\subfigure[Forecasting Accuracy on Highly Seasonal Time Series.]{\vspace{-0.4in}
	\label{fig:high_seasonality_results}\includegraphics[scale=0.33]{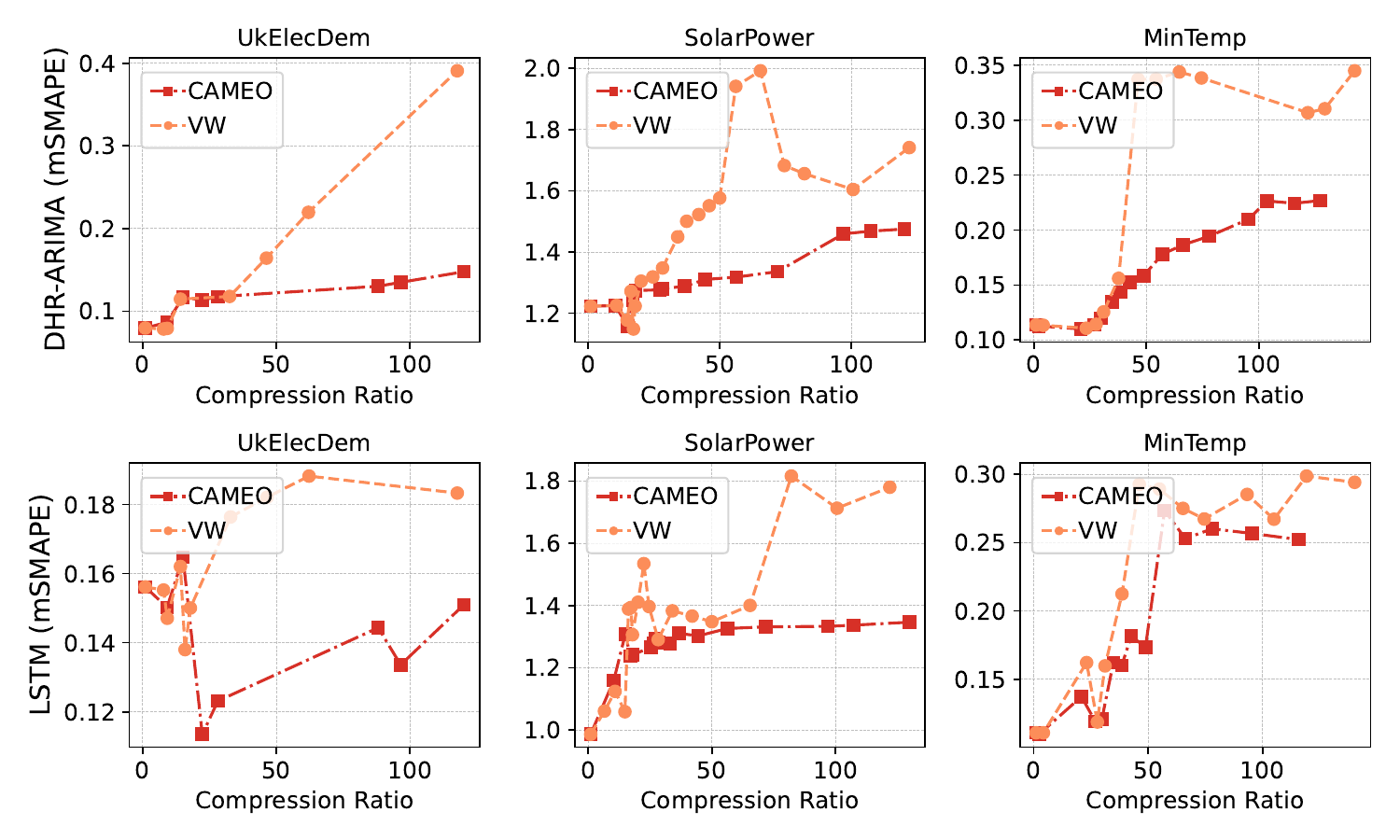}\vspace{-0.4in}}
\vspace{-0.1in}
\caption{\label{fig:combined_results}Impact on Forecasting Accuracy of CAMEO and Baselines as the Compression Ratio Increases under Different Configuration, Quality Metrics, and Time Series.}
\vspace{-0.1in}
\end{figure}

\textbf{Distance Metrics Evaluation (EXP1):} We test CAMEO using RMSE, MAPE, Chebyshev Distance (CHEB)~\cite{cantrell2000modern}, and MAE against the rest of line-simplification baselines (VW, TP, and PIP). We use the Pedestrian dataset~\cite{DBLP:conf/nips/GodahewaBWHM21}, comprising 66 time series of varying lengths and statistical properties. We segment them into chunks of \numprint{10000} points to ensure consistency. We then apply a Box-Cox~\cite{box1964analysis} power transformation to stabilize variance and improve normality, followed by standardization to ensure a common scale across all series. We switch to the compression-centric (Definition~\ref{def:data-centric}) approach to control the CR from 2 (half the points sampled) to 10 (only 1000 points sampled) for CAMEO and the baselines. This process generated a total of \numprint{3400} series on which we conduct training and forecast the last 24 points using the Holt-Winters model~\cite{chatfield1988holt}.
Figure~\ref{fig:distant_metrics} shows the results using MSE and MAPE as quality metrics. The results show that CAMEO consistently maintains forecasting accuracy longer than the line simplification baselines. Among CAMEO's variants, MAPE performed the worst, while CHEB was the best, showing almost no accuracy loss until a CR of 6. Intuitively, MAPE focuses on preserving low-value regions, while CHEB distributes the ACF error evenly, avoiding distortion on specific lags, which aggregating metrics can obscure. 

\textbf{Lossy Compression Evaluation (EXP2):} This time we follow Godahewa's benchmark~\cite{DBLP:conf/nips/GodahewaBWHM21} on the Pedestrian dataset which uses directly the 66 times series no extra preprocessing and forecasts the last 24 points using STL-ARIMA, STL-ETS, and LSTM~\cite{DBLP:journals/neco/GersSC00} forecasting models. All models are trained on the compressed data, and their forecasting accuracy is evaluated against the raw data. We perform a trial-and-error exploration of the error bound of the lossy compressor as we cannot control the compression error beforehand. Figure~\ref{fig:forecasting_impact} shows the results using mSMAPE. The results show that CAMEO can generally preserve the average forecasting accuracy equal to or better than the baselines at various compression ratios. Particularly, when using LSTM, CAMEO maintains, and even slightly improves, the mSMAPE until 10x of compression. VW also yields very good results, especially using STLF-ARIMA. The variable results across forecasting models can be attributed to the different time series characteristics, especially seasonality measured by the seasonal strength~\cite{wang2006characteristic}.

\textbf{Highly Seasonal Data Evaluation (EXP3):} We evaluate CAMEO's and VW's impact on forecasting accuracy on highly seasonal time series. We use UKElecDem, SolarPower, and MinTemp, which all have a high seasonal strength of 0.84, 0.81, and 0.6, respectively. Figure~\ref{fig:high_seasonality_results} shows a segment of the three datasets and the forecasting accuracy (mSMAPE) as the compression ratio increases for two models: Dynamic Harmonic Regression (DHR) ARIMA~\cite{young1999dynamic, hyndman2018forecasting} and LSTM. The remarkable results support our hypothesis that preserving the ACF is beneficial for forecasting accuracy when applying lossy compression. Notably, even when the compression ratio approaches 100x, CAMEO steadily preserves the forecasting accuracy. We attribute these results to the strong seasonality in these three datasets and CAMEO's effective selection of a few data points capable of preserving their seasonality. 

\textbf{Forecasting Experiments Conclusion:} These three experiments show CAMEO's superiority over the baselines across datasets, forecasting models, and evaluation metrics, making a strong case for its use in compression scenarios involving forecasting as downstream analytics.

\vspace{-0.05in}
\subsection{Impact on Anomaly Detection}
\label{sec:anomaly}
In a last set of experiments, we investigate two alternative hypotheses: (1) preserving the ACF during compression is beneficial for anomaly detection, and (2) CAMEO execution time is amortized if the downstream analytics can exploit the resulting (much smaller) irregular time series. For the first hypothesis, we use the UCR dataset~\cite{DBLP:conf/icde/WuK22} consisting of 250 time series and the Matrix Profile (MP) algorithm~\cite{DBLP:conf/icdm/YehZUBDDSMK16}. We measure the accuracy using the UCR-score~\cite{DBLP:conf/icde/WuK22}, where higher scores indicate better detection. We detect all discords using the MP algorithm with segment sizes ranging from 75 to 125 and select the one with the maximum distance~\cite{is_it_worth_it}. For the second hypothesis, we implement an algorithm that calculates the Euclidean distance between all pairs of segments of size $m$--- MP's core idea---over the irregular time series (iMP). 
iMP avoids materializing the data and directly computes distances by leveraging linear interpolation during decompression using the remaining $m'$ points per segment. This method reduces the complexity from $\mathcal{O}(N^2m)$---the complexity of naive implementation over regular time series (rMP)---to $\mathcal{O}(N^2m')$, where $ m' \ll m$ and $N$ is the time series length. We conduct tests on synthetically generated data of size $2^{p}$, where $p$ ranges from $10$ to $16$, and segment size $m=150$.

\begin{figure}[!t]
\includegraphics[scale=1, height=2.5cm]{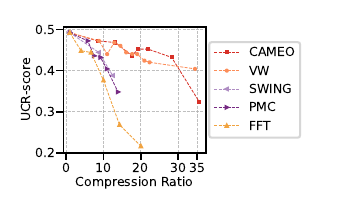} 
\hfil
\includegraphics[scale=.97, height=2.5cm]{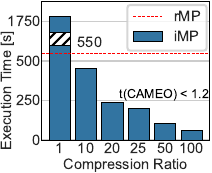} 
\vspace{-0.25cm}
\caption{{\normalfont(left)} Impact on the Anomaly Detection Accuracy as the Compression Ratio Increases. {\normalfont(right)} Execution Time of the MP Algorithm over the Irregular Time Series.}
\label{fig:anomaly_detection_impact}
\end{figure}

\textbf{Accuracy Results Analysis:} Figure~\ref{fig:anomaly_detection_impact} (left) illustrates the UCR-score as the compression ratio increases. The results show that CAMEO preserves the UCR-score more effectively than PMC, SWING, and FFT, achieving a compression ratio of $\thickapprox 28$x while minimally impacting accuracy. The results support our hypothesis (1) that preserving the ACF is advantageous for forecasting analytics and other applications such as anomaly detection. However, it is noteworthy that the effectiveness of preserving the ACF diminishes at higher compression ratios, as shown in Figure~\ref{fig:anomaly_detection_impact} (left) beyond a 30x ratio. This is likely because removing extreme outliers has a negligible impact on the ACF since these points do not significantly affect temporal dependencies. Thus, removing these points negatively affects the Euclidean distance computation. In contrast, the VW strategy implicitly retains such points, as an outlier typically has a significant triangular area.

\textbf{Execution Time Results Analysis:} Figure~\ref{fig:anomaly_detection_impact} (right) displays the execution time results for iMP as the compression ratio increases, specifically for $p=14$. The results reveal a significant reduction in execution time, decreasing from 550 seconds with the naive implementation rMP to 250 at a compression ratio of 20x. Furthermore, the compression process with CAMEO is negligible, requiring only 0.94 seconds to complete at that compression ratio and less than 1.2 seconds at a ratio of 100x. Experiments for different $p$ values show similar results. This improvement of the end-to-end runtime of the analytics, coupled with the minimal impact on detection accuracy, highlights the substantial benefits of using compression algorithms like CAMEO that preserve key statistical features while significantly reducing the data size.

\section{Additional Related Work}

Here we position CAMEO in the context of additional work, including line simplification, lossless time series, and matrix compression.

\textbf{Line Simplification:} Algorithms based on line simplification---also known as curve simplification---are used to reduce the number of points of a given curve composed of line segments, also called polyline, while preserving its global shape~\cite{rammer1972iterative,douglas1973algorithms,visvalingam1993line,kronenfeld2020simplification,raposo2013scale}. Many existing algorithms follow an iterative procedure to remove the points while minimizing the areal displacement, preserving the polygonal areas, or a combination of both~\cite{kronenfeld2020simplification}. One of the earliest effective algorithms for line simplification is Ramer-Douglas-Peucker (RDP)~\cite{douglas1973algorithms, rammer1972iterative}. RDP retains the points that lie further than a given tolerance from the lines connecting endpoints, emphasizing geometric accuracy. In contrast, VW~\cite{visvalingam1993line} focuses on visual quality, removing points that contribute least to the perceived shape of the polyline. Unlike both methods, CAMEO preserves complex statistical features that are aggregates of the entire time series.

\textbf{Lossless Time Series Compression:}  Work on lossless time series compression has witnessed increasing improvements and nowadays yields a good balance between compression ratio and computational efficiency~\cite{blalock2018sprintz, campobello2017rake, DBLP:conf/dcc/RatanaworabhanKB06}. Gorilla~\cite{pelkonen2015gorilla}, widely known for its implementation within Facebook's time-series database, is simple yet very efficient, making it amenable for real-time applications. Gorilla's XOR operator has recently inspired the Chimp~\cite{liakos2022chimp} and Elf~\cite{10.14778/3587136.3587149} lossless compression algorithms. Both methods preserve Gorilla's linear time complexity while improving its compression ratio for time series without many repeating values. 
However, as shown in Section~\ref{sec:experiments}, the compression ratios of these lossless methods are still very limited. Instead, CAMEO is positioned between lossless and lossy compression by preserving key statistical features while yielding very good compression ratios.

\textbf{Lossless Matrix and Workload-Aware Compression:} Besides lossy matrix compression---which are mainstream in ML model training and inference---there is also some work on lossless matrix compression. Examples are compressed linear algebra~\cite{DBLP:journals/vldb/ElgoharyBHRR18, elgohary2016compressed} and tuple-oriented coding~\cite{DBLP:conf/sigmod/LiCZ00NP19}, which also apply to time series data but only in combination with binning or quantization. Recent work also explored workload-aware lossless compression~\cite{DBLP:journals/pacmmod/Baunsgaard023} and workload-aware dimensionality reduction \cite{DBLP:conf/sigmod/SuriB19}, which are related to our compression for downstream analytics. Workload-aware compression can combine lossy and lossless compression, offering a strategy where data compression is driven not purely by fidelity (as in lossless compression) or storage (as in lossy compression), but also by the analytics that use the data. A holistic, unified strategy that dynamically applies the principles of lossless compression, lossy compression, and workload-aware compression (especially for statistical features) is, however, non-existent so far.

\section{Conclusions}

We introduced CAMEO, a lossy time series compression framework that guarantees a user-provided maximum deviation of the original ACF/PACF. Inspired by line simplification methods, CAMEO iteratively removes points while continuously validating the error constraint. To improve efficiency, CAMEO utilizes incremental maintenance of statistical properties, blocking, and parallelization strategies. 
Based on our experimental evaluation, we draw the following conclusions: 1) CAMEO obtains higher compression ratios than existing line-simplification techniques while preserving the same ACF deviation. 2) CAMEO provides a competitive alternative to the well-known lossy compressors PMC, SWING, and SP with a better compression ratio and direct guarantees on the ACF. 3) Preserving the ACF during compression yields better forecasting accuracy across different time series and forecasting models. 
Together, these promising results make a great case for lossy compression under awareness of statistical properties and downstream applications, which helps to remove trust concerns on lossy compression and tedious semi-manual trial-and-error exploration. 

\begin{acks}
This paper was partly supported by the MORE project funded by the EU Horizon 2020 program under grant agreement no. 957345. Additionally, we acknowledge funding from the German Federal Ministry of Education and Research (under research grant BIFOLD24B).
\end{acks}

\balance
\bibliographystyle{ACM-Reference-Format}
\bibliography{cameo}

\end{document}